\documentstyle[12pt]{article}
\newcommand{\be}{\begin{equation}}
\newcommand{\ee}{\end{equation}}
\newcommand{\rarrow}{\rightarrow}

\begin{document}
\begin{titlepage}
\hspace{11.3cm}{\small{CALT-68-1960}}

\hspace{11.3cm}{\small{October 1994}}
\vspace*{2cm}
\begin{center}
{\Large \bf On the Classification of Fusion Rings}\\
\vskip50pt \bf Doron Gepner \footnote[1]{On leave of absence from
 Weizmann Institute, Israel} \footnote[2]{Work supported in part by the
 U.S.Dept. of Energy under Grant No. DE-FG03-92-ER40701} \\
\vskip14pt \bf Anton Kapustin $^\dagger$ \\
\vskip14pt {\it Department of Physics, Mathematics and Astronomy \\
California Institute of Technology \\ Pasadena, CA 91125}\\
\vskip40pt
\abstract{
The fusion rules and modular matrix of a rational conformal field theory
obey a list of properties. We use these properties to classify rational
conformal field theories with not more than six primary fields
and small values of the fusion coefficients.
We give a catalogue of fusion rings which can arise for these field theories.
It is shown
that all such fusion rules can be realized by current algebras. Our results
 support the conjecture that all rational conformal field theories  are
related to current algebras.}

\end{center}
\end{titlepage}
\baselineskip=17pt
Rational conformal field theories have attracted considerable attention
starting with the seminal work of BPZ \cite{BPZ}. One main question is the
problem of
the classification of rational conformal field theories (RCFT). It was
suggested \cite{MooSei} that all such theories stem from
current algebras via the coset construction or various extended algebras.
The first step is to try to establish if all fusion rings of RCFT stem from
current algebras. This question was discussed in \cite{Cas1,Cas2,Eho}; in
 particular in \cite{Cas1} it was shown that all fusion rings with three or
less
primaries originate from current algebras.
In this note we wish to address this question by making a computer
classification of fusion rings of RCFT for low number of primary fields and a
 low value for the fusion coefficients.

Our results can be summarized as follows. We studied RCFT with the number
of primary fields $n\leq6$ and the fusion coefficients $N_{ij}^k\leq
 100,10,3,1$ for $n=3,4,5,6$ respectively.
We find that all the theories found in this range can be obtained from
current algebras, and that no ``exotic'' theories exist.

Our starting point is the fusion rules and the modular transformations of
the RCFT, following the connection found in ref. \cite{Verlinde}, between the
two. In an RCFT these are required to obey the
following set of properties:

1) The fusion rules form an associative commutative ring. The structure
constants are non--negative integers, $N_{ij}^k$, which obey
$N_{ij}^k N_{kl}^m=N_{il}^k N_{kj}^m$.

2) There is a symmetric bilinear form $b_{ij}$ such that $b_{ij}=0$,
except for $i=\bar j$, where $\bar j$ is some permutation of $j$, and
where $b_{i,\bar i}=1$. The structure constant $N_{ijk}=b_{im} N^m_{jk}$,
is symmetric in all three indices. $N^0_{ij}=b_{ij}$, where $0$ denotes
the identity field.

3) The matrix $S$ of modular transformations is a symmetric unitary matrix
whose order is the number of primary fields and is related to the fusion
rules by
\be
N^k_{ij}=\sum_m {S_{im} S_{jm} S^\dagger_{km}\over S_{0m}}. \label{Verlf}
\ee
The matrix $S$ obeys $(S^2)_{ij}=b_{ij}$, and  implements the modular
transformation $\tau\rarrow{-1\over\tau}$. In addition, the
diagonal matrix $T=\exp[2\pi i(\Delta_p-c/24)]$, where $\Delta_p$ is the
dimension of the primary field $p$ and $c$ is the central charge,
implements the modular
transformation $\tau\rarrow\tau+1$ and obeys $(ST)^3=b$.

4) The structure constants and the dimensions obey the equations \cite{Vafa}
\be
(\alpha_i\alpha_j\alpha_k\alpha_l)^{N_{ijkl}}=\prod_r \alpha_r^{N_{ijkl,r}},
\label{Vafaeq}
\ee
where,
\be
\alpha_r=\exp(2\pi i\Delta_r),\qquad N_{ijkl}=N_{ij}^{\bar m} N_{kl}^m,\qquad
N_{ijkl,r}=N_{ij}^r N_{klr}+N_{jk}^r N_{ilr}+N_{ik}^r N_{jlr}. \label{defVafa}
\ee
An RCFT obeys additional properties such as the braid group representation
\cite{Braid}. However, on the basis of our study of theories with
low number of primary fields we suggest that the properties (1-4) fully
characterize an RCFT, and that any commutative ring satisfying these
properties is the fusion ring of some full fledged RCFT. The question is thus
to classify rings obeying the properties (1-4).

This problem is very difficult to settle completely. We thus resort to a
computer study which
can be readily done for a small number of primary fields and a small value
for the fusion coefficients $N_{ijk}$.
The following procedure was employed to generate fusion rings. We select
one primary field, denoted by `$1$', different from the identity.
We than fix the structure constants $N_{1j}^k$, subject to the condition
that $N_{1jk}$ is fully symmetric. We can always find such values for the rest
of the fusion coefficients that the fusion ring be associative.
 Now, in the vast majority of cases, $N_{1j}^k$
determines the full fusion ring, by associativity. Degenerate cases, where
this is not so, are treated separately. We run the program over
all values of $N_{1i}^j$ consistent with the symmetry of $N_{1ij}$ and
for a fixed small number of primary fields and some upper bound for $N_{1i}^j
\leq M$, where $M$ is small.

For each selection we proceed to construct the $S$ matrix by solving the
system of equations,
\be
\psi_1 \psi_j=N_{1j}^k \psi_k,\label{ringrep}
\ee
where $\psi_i$ are some complex numbers. This is essentially a linear problem,
whose eigenvectors span the $n$-dimensional space where $n$ is the number of
 primary fields, if $N_{1i}^j$ is a multiplication matrix of a fusion ring. The
 solutions $\psi_i$ are related by property (3) to the $S$ matrix by
\be
\psi_i^{(j)}={S_{ij}\over S_{i0}}. \label{formulaforS}
\ee
We thus construct the matrix $S$ by taking the matrix $\psi_i^{(j)}$, permuting
arbitrarily its columns, and normalizing each column to have  norm $1$,
up to an undetermined $\pm 1$ sign ($S_{0j}$ must be all real).
 We then check if the matrix $S$ is
symmetric (it is automatically unitary). This is a very stringent condition
which rules out all but a very small number of rings. For these we solve
the equations (\ref{Vafaeq}) to find the possible dimensions, and check for
each solution
if $(ST)^3=b$ is satisfied. This concludes the search for rings satisfying
(1-4), except for the degenerate cases. For these we generate the remaining
fusion coefficients, check for associativity and symmetry and proceed in the
same way.

We present our results in tables (1-5). These tables are a complete catalogue
of fusion rings with not more than six primary fields (with the
above mentioned limitations
on the structure constants). All the fusion rings that we find can be realized
in terms of current algebras. We find theories of two types: 1) WZW models
\cite{WZW}; 2) Theories obtained by taking the extended
algebra of products of current algebras \cite{Found}.
The fusion rings of WZW theories were described in ref. \cite{GW}.
The fusion rings of type 2 are closely related to the latter
 and can be described,
for our puposes, as the sub--rings of representations invariant under the
center
of the corresponding WZW model. For example, $SU(3)_4/Z_3$ which has five
primary fields has a fusion ring which is the same as $SU(3)_4$ WZW theory,
but with only $Z_3$ (color) singlet representations retained.
 It is remarkable that one  can always find realizations for fusion rings with
six or less primaries  in terms of the RCFTs of these two types. Note that
other
 types of RCFT are known, such as orbifolds
\cite{Orbi} and coset models \cite{Coset}.
However, these  do not seem to be required for the realizations. It is an
interesting question whether this situation persists for more that six
primaries.

Note that the same fusion rules correspond, in general, to a number of
different
modular matrices. In the context of WZW theories, this ambiguity was described
in ref \cite{Found}. The different $S$ matrices are obtained from each other by
some permutation of the columns. It is presently unkown whether all such $S$
matrices are actually realized as modular matrices of some RCFT. For this
reason, we cannot list realizations for all the modular matrices that we
find.

This brings us to the question if all RCFT can be constructed from current
algebras by cosets, orbifolds and so forth. In other words, do current
algebras completely classify RCFT? We investigate here a slightly weaker
version of this question: are all fusion rings realized by current algebras.
Our findings indicate that this is indeed the case. This lends considerable
support to the conjecture that all RCFT can be realized in terms of current
 algebras, and that we essentially know already all RCFT.
This is a somewhat disappointing finding. We know that the set of non-rational
conformal field theories is very rich, including moduli. It is unfortunate
that only current algebras, then, have the vast simplification and beauty
of rationality.

Our investigation was carried only up to six primary fields. However, a
stronger computer than the desk--top we used can be utilized to carry the
search to a higher number of primary fields without any change in the
algorythm. We believe that investigation up to ten or so primary fields is
quite feasible. We strongly suspect, however, that the main result will
persist, and that all fusion rings can be realized in terms of current
 algebras.

Finally, we hope that this work helps to clarify a question that has been
in the air for quite some time. Namely, what are the RCFT and do we
already know all of them. In view of the many connections of RCFT to
integrable models, solvable lattice systems, etc., this question is
of particular interest.\newline\newline
{\bf Acknowledgements.} We would like to thank S. Cherkis for numerous
 discussions.

\newpage


\newcommand{\f}{\phi}


\begin{table}[htb]
\begin{tabular}{c|c|c|c}
& Fusion ring & Central charge $mod4$ & Realization \\ \hline\hline
 & $\f^2=1$    & $ 1,3 $                        & $SU(2)_1$ \\ \hline
 & $\f^2=1+\f$ & $\frac{4n+2}{5}, n<5, n\neq 2$ & $SU(2)_3/Z_2 $ \\ \hline
\end{tabular}
\caption{Fusion rings with two primaries.}
\end{table}


\begin{table}[h]
\begin{tabular}{c|c|c|c}
& Fusion ring & Central charge $mod4$ & Realization \\ \hline\hline

 &
\begin{tabular}{cc}
$\f_1^2=1$, & $\f_2^2=1+\f_1$, \\
            & $\f_1\f_2=\f_2$
\end{tabular}
 & $\frac{2n+1}{2}, n<4$ & $SU(2)_2$ \\ \hline

 &
\begin{tabular}{cc}
 $\f_1^2=1+\f_2$, & $\f_2^2=1+\f_1+\f_2$, \\
                  & $\f_1\f_2=\f_1+\f_2$
\end{tabular}
& $\frac{4n}{7}, n<7$ & $SU(2)_5/Z_2$ \\ \hline

 &
\begin{tabular}{cc}
 $\f_1^2=\f_2$, & $\f_2^2=\f_1$, \\
                & $\f_1\f_2=1$ \\
\end{tabular}
 & $2$ & $SU(3)_1$ \\ \hline
\end{tabular}
\caption{Fusion rings with three primaries and fusion coefficients $N_{ijk}
\leq 100$.}
\end{table}


\begin{table}[htb]
\begin{tabular}{c|c|c|c}
& Fusion ring &
\begin{tabular}{c}
Central charge \\
$mod4$
\end{tabular}
& Realization \\ \hline\hline

&
\begin{tabular}{c}
 $\f_1^2=1, \f_2^2=1$, \\
 $\f_3^2=1, \f_1\f_2=\f_3$, \\
 $\f_1\f_3=\f_2, \f_2\f_3=\f_1$
\end{tabular}
& $ 0,2 $ & \begin{tabular}{c} $SU(2)_1\times SU(2)_1$ \\
 or $SO(8)_1$ \end{tabular} \\ \hline

&
\begin{tabular}{c}
 $\f_1^2=1, \f_2^2=1+\f_2$, \\
 $\f_3^2=1+\f_2$, \\
 $\f_1\f_2=\f_3, \f_1\f_3=\f_2$, \\
 $\f_2\f_3=\f_1+\f_3$
\end{tabular}
 &
\begin{tabular}{c}
$\frac{2n+1}{5}$, \\
$ n<10$,           \\
$n\neq 2,7$
\end{tabular}
& $SU(2)_3$ \\ \hline

&
\begin{tabular}{c}
 $\f_1^2=1+\f_1$, \\
 $\f_2^2=1+\f_2$, \\
 $\f_3^2=1+\f_1+\f_2+\f_3$, \\
 $\f_1\f_2=\f_3$, \\
 $\f_1\f_3=\f_2+\f_3$, \\
 $\f_2\f_3=\f_1+\f_3$
\end{tabular}
 &
\begin{tabular}{c}
$\frac{2n}{5}$, \\
$n<9$,          \\
$n\neq 3,5,7$
\end{tabular}
& $SU(2)_3/Z_2\times SU(2)_3/Z_2$ \\ \hline

&
\begin{tabular}{c}
 $\f_1^2=1+\f_2$, \\
 $\f_2^2=1+\f_2+\f_3$, \\
 $\f_3^2=1+\f_1+\f_2+\f_3$, \\
 $\f_1\f_2=\f_1+\f_3$, \\
 $\f_1\f_3=\f_2+\f_3$, \\
 $\f_2\f_3=\f_1+\f_2+\f_3$
\end{tabular}
& $\frac{2}{3},\frac{10}{3}$ & $SU(2)_7/Z_2$ \\ \hline

&
\begin{tabular}{c}
 $\f_1^2=1, \f_2^2=\f_1$, \\
 $\f_3^2=\f_1, \f_1\f_2=\f_3$, \\
 $\f_1\f_3=\f_2, \f_2\f_3=1$
\end{tabular}
& $1,3$ & $SU(4)_1$ \\ \hline
\end{tabular}
\caption{Fusion rings with four primaries and $N_{ijk}\leq 10$.}
\end{table}

\begin{table}[htb]
\begin{tabular}{c|c|c|c}
& Fusion ring &
\begin{tabular}{c}
Central charge \\
$mod4$
\end{tabular}
& Realization \\ \hline\hline

&
\begin{tabular}{c}
$\f_1^2=1+\f_2$, \\
$\f_2^2=1+\f_2+\f_4$, \\
$\f_3^2=1+\f_2+\f_3+\f_4$, \\
$\f_4^2=1+\f_1+\f_2+\f_3+\f_4$, \\
$\f_1\f_2=\f_1+\f_3$, \\
$\f_1\f_3=\f_2+\f_4$, \\
$\f_1\f_4=\f_3+\f_4$, \\
$\f_2\f_3=\f_1+\f_3+\f_4$, \\
$\f_2\f_4=\f_2+\f_3+\f_4$, \\
$\f_3\f_4=\f_1+\f_2+\f_3+\f_4$
\end{tabular}
&
 \begin{tabular}{c}
$\frac{4(n+1)}{11}$, \\
$n<9$
\end{tabular}
& $SU(2)_9/Z_2$ \\ \hline

&
\begin{tabular}{c}
$\f_1^2=1+\f_2$, \\
$\f_2^2=1+\f_2+\f_4$, \\
$\f_3^2=1+\f_2, \f_4^2=1$, \\
$\f_1\f_2=\f_1+\f_3$, \\
$\f_1\f_3=\f_2+\f_4$, \\
$\f_1\f_4=\f_3, \f_2\f_3=\f_1+\f_3$, \\
$\f_2\f_4=\f_2, \f_3\f_4=\f_1$
\end{tabular}
& $2$ & $SU(2)_4$ \\ \hline

&
\begin{tabular}{c}
$\f_1^2=\f_3, \f_2^2=\f_4$, \\
$\f_3^2=\f_2, \f_4^2=\f_1$, \\
$\f_1\f_2=1, \f_1\f_3=\f_4$, \\
$\f_1\f_4=\f_2, \f_2\f_3=\f_1$, \\
$\f_2\f_4=\f_3, \f_3\f_4=1$
\end{tabular}
& $0$ & $SU(5)_1$ \\ \hline

&
\begin{tabular}{c}
$\f_1^2=1+\f_1+\f_2$, \\
$\f_2^2=1+\f_1+2\f_2+\f_3+\f_4$, \\
$\f_3^2=\f_1+\f_4$, \\
$\f_4^2=\f_1+\f_3$, \\
$\f_1\f_2=\f_1+\f_2+\f_3+\f_4$, \\
$\f_1\f_3=\f_2+\f_4$, \\
$\f_1\f_4=\f_2+\f_3$, \\
$\f_2\f_3=\f_1+\f_2+\f_3$, \\
$\f_2\f_4=\f_1+\f_2+\f_4$, \\
$\f_3\f_4=1+\f_2$
\end{tabular}
&
\begin{tabular}{c}
$\frac{4n+2}{7}$, \\
$n<7$, \\
$n\neq 3$
\end{tabular}
& $SU(3)_4/Z_3$ \\ \hline
\end{tabular}
\caption{Fusion rings with five primaries and $N_{ijk}\leq 3$.}
\end{table}


\begin{table}[htb]
\begin{tabular}{c|c|c|c}
& Fusion ring &
\begin{tabular}{c}
Central charge \\
$mod4$
\end{tabular}
& Realization \\ \hline\hline

&
\begin{tabular}{c}
$\f_1^2=1, \f_2^2=2$, \\
$\f_3^2=1+\f_2, \f_4^2=1$, \\
$\f_5^2=1+\f_2, \f_1\f_2=\f_4$, \\
$\f_1\f_3=\f_5, \f_1\f_4=\f_2$, \\
$\f_1\f_5=\f_3, \f_2\f_3=\f_3$, \\
$\f_2\f_4=\f_1, \f_2\f_5=\f_5$, \\
$\f_3\f_4=\f_5, \f_3\f_5=\f_1+\f_4$, \\
$\f_4\f_5=\f_3$
\end{tabular}
& \begin{tabular}{c} $\frac{2n+1}{2}$, \\ $n<4$ \end{tabular}
& $SU(2)_2\times SU(2)_1$ \\ \hline

&
\begin{tabular}{c}
$\f_1^2=1, \f_2^2=1+\f_3$, \\
$\f_3^2=1+\f_2+\f_3$, \\
$\f_4^2=1+\f_3$, \\
$\f_5^2=1+\f_2+\f_3$, \\
$\f_1\f_2=\f_4, \f_1\f_3=\f_5$, \\
$\f_1\f_4=\f_2, \f_1\f_5=\f_3$, \\
$\f_2\f_3=\f_2+\f_3$, \\
$\f_2\f_4=\f_1+\f_5$, \\
$\f_2\f_5=\f_4+\f_5$, \\
$\f_3\f_4=\f_4+\f_5$, \\
$\f_3\f_5=\f_1+\f_4+\f_5$, \\
$\f_4\f_5=\f_2+\f_3$
\end{tabular}
& \begin{tabular}{c}
$\frac{2n+1}{7}$, \\
$n<14$, \\
$n\neq 3,10$
\end{tabular}
& $SU(2)_5$ \\ \hline

&
\begin{tabular}{c}
$\f_1^2=1, \f_2^2=\f_3$, \\
$\f_3^2=\f_2, \f_4^2=\f_3$, \\
$\f_5^2=\f_2, \f_1\f_2=\f_4$, \\
$\f_1\f_3=\f_5, \f_1\f_4=\f_2$, \\
$\f_1\f_5=\f_3, \f_2\f_3=1$, \\
$\f_2\f_4=\f_5, \f_2\f_5=\f_1$, \\
$\f_3\f_4=\f_1, \f_3\f_5=\f_4$, \\
$\f_4\f_5=1$
\end{tabular}
& $1,3$ & $SU(6)_1$ \\ \hline
\end{tabular}
\caption{Fusion rings with six primaries and $N_{ijk}\leq 1$.}
\end{table}


\begin{table}[htb]
\begin{tabular}{c|c|c|c}
& Fusion ring &
\begin{tabular}{c}
Central \\
charge $mod4$
\end{tabular}
& Realization \\ \hline\hline

&
\begin{tabular}{c}
$\f_1^2=1+\f_1, \f_2^2=1+\f_3$, \\
$\f_3^2=1+\f_2+\f_3$, \\
$\f_4^2=1+\f_1+\f_3+\f_5$, \\
$\f_5^2=1+\f_1+\f_2+\f_3$ \\
$+\f_4+\f_5$, \\
$\f_1\f_2=\f_4, \f_1\f_3=\f_5$, \\
$\f_1\f_4=\f_2+\f_4$, \\
$\f_1\f_5=\f_3+\f_5$, \\
$\f_2\f_3=\f_2+\f_3$, \\
$\f_2\f_4=\f_1+\f_5$, \\
$\f_2\f_5=\f_4+\f_5$, \\
$\f_3\f_4=\f_4+\f_5$, \\
$\f_3\f_5=\f_1+\f_4+\f_5$, \\
$\f_4\f_5=\f_2+\f_3$ \\
$+\f_4+\f_5$
\end{tabular}
&
\begin{tabular}{c}
$\frac{4n+2}{35}$, \\
$n<35$, \\
$n\neq 2,3,7,10$, \\
$12,22,24$, \\
$27,31,32$
\end{tabular}
& \begin{tabular}{c} $SU(2)_3/Z_2$ \\
$\times SU(2)_5/Z_2$ \end{tabular} \\ \hline

&
\begin{tabular}{c}
$\f_1^2=1+\f_1, \f_2^2=1$, \\
$\f_3^2=1+\f_2, \f_4^2=1+\f_1$, \\
$\f_5^2=1+\f_1+\f_2+\f_4$, \\
$\f_1\f_2=\f_4, \f_1\f_3=\f_5$, \\
$\f_1\f_4=\f_2+\f_4$, \\
$\f_1\f_5=\f_3+\f_5$, \\
$\f_2\f_3=\f_3, \f_2\f_4=\f_1$, \\
$\f_2\f_5=\f_5, \f_3\f_4=\f_5$, \\
$\f_3\f_5=\f_1+\f_4$, \\
$\f_4\f_5=\f_3+\f_5$
\end{tabular}
& \begin{tabular}{c}  $\frac{2n+1}{10}$, \\ $n<20$, \\ $n\neq 2,7,12,17$
\end{tabular}
& \begin{tabular}{c} $SU(2)_3/Z_2$ \\
$\times SU(2)_2$ \end{tabular} \\ \hline

&
\begin{tabular}{c}
$\f_1^2=1+\f_1, \f_2^2=\f_3$, \\
$\f_3^2=\f_2, \f_4^2=\f_3+\f_5$, \\
$\f_5^2=\f_2+\f_4$, \\
$\f_1\f_2=\f_4, \f_1\f_3=\f_5$, \\
$\f_1\f_4=\f_2+\f_4$, \\
$\f_1\f_5=\f_3+\f_5$, \\
$\f_2\f_3=1, \f_2\f_4=\f_5$, \\
$\f_2\f_5=\f_1, \f_3\f_4=\f_1$, \\
$\f_3\f_5=\f_4, \f_4\f_5=1+\f_1$
\end{tabular}
& \begin{tabular}{c} $\frac{4n+4}{5}$, \\ $n<5$ \end{tabular} & $SU(3)_2$ \\
\hline
\end{tabular}
\end{table}


\begin{table}[htb]
\begin{tabular}{c|c|c|c}
& Fusion ring &
\begin{tabular}{c}
Central \\
charge $mod4$
\end{tabular}
& Realization \\ \hline\hline

&
\begin{tabular}{c}
$\f_1^2=1+\f_2$, \\
$\f_2^2=1+\f_2+\f_4$, \\
$\f_3^2=1+\f_2+\f_4+\f_5$, \\
$\f_4^2=1+\f_2+\f_3$ \\
$+\f_4+\f_5$, \\
$\f_5^2=1+\f_1+\f_2$, \\
$+\f_3+\f_4+\f_5$, \\
$\f_1\f_2=\f_1+\f_3$, \\
$\f_1\f_3=\f_2+\f_4$, \\
$\f_1\f_4=\f_3+\f_5$, \\
$\f_1\f_5=\f_4+\f_5$, \\
$\f_2\f_3=\f_1+\f_3+\f_5$, \\
$\f_2\f_4=\f_2+\f_4+\f_5$, \\
$\f_2\f_5=\f_3+\f_4+\f_5$, \\
$\f_3\f_4=\f_1+\f_3+\f_4+\f_5$, \\
$\f_3\f_5=\f_2+\f_3+\f_4+\f_5$, \\
$\f_4\f_5=\f_1+\f_2+\f_3$, \\
$+\f_4+\f_5$
\end{tabular}
& \begin{tabular}{c} $\frac{4n+2}{13}$, \\ $n<13$, \\ $n\neq 6$ \end{tabular}
& $SU(2)_{11}/Z_2$ \\ \hline

&
\begin{tabular}{c}
$\f_1^2=1$, \\
$\f_2^2=1+\f_3+\f_4$, \\
$\f_3^2=1+\f_1+\f_4$, \\
$\f_4^2=1+\f_1+\f_3$, \\
$\f_5^2=1+\f_3+\f_4$, \\
$\f_1\f_2=\f_5, \f_1\f_3=\f_3$, \\
$\f_1\f_4=\f_4, \f_1\f_5=\f_2$, \\
$\f_2\f_3=\f_2+\f_5$, \\
$\f_2\f_4=\f_2+\f_5$, \\
$\f_2\f_5=\f_1+\f_3+\f_4$, \\
$\f_3\f_4=\f_3+\f_4$, \\
$\f_3\f_5=\f_2+\f_5$, \\
$\f_4\f_5=\f_2+\f_5$
\end{tabular}
& $0$ & $SO(5)_2$ \\ \hline
\end{tabular}
\end{table}
\end{document}